\documentclass[12pt]{article}        %%%%%%%%%%%%%%%%%%
\usepackage{color}

\usepackage{amsmath} 
\usepackage{amssymb} 
\usepackage{euscript}

\usepackage{epsfig}

\usepackage{cite}

\usepackage{alltt}

\usepackage{epic}
\begin{document}                     %%%%%%%%%%%%%%%%%%%%%%

\allowdisplaybreaks

%%%%%%%%%%%%%%%%%%%%%%%%%%%%%%%%%%%%%%%%%%%%%%%%%%%%%%%%%%%%%%%%%%%%%%%%%%%%%%%%%
%%%%%%%%%%%%%%%%%%%%%%%%%%%%%%%%%%%%%%%%%%%%%%%%%%%%%%%%%%%%%%%%%%%%%%%%%%%%%%%%%
\begin{titlepage}

\begin{flushright}
 {\bf CERN-TH/2003-287} \\  {\bf December 2003}
\end{flushright}
\vspace{0.6 cm}

\begin{center}{\bf\Large The tauola-photos-F environment }\end{center}
\begin{center}{\bf\Large for  the TAUOLA and PHOTOS packages$^{\dag}$,}\end{center}
\begin{center}{\bf\Large \em
  release II}\end{center}
\vspace{0.8 cm}
\begin{center}
{\large \bf P. Golonka$^{a,b,c}$,} 
{\large \bf B. Kersevan$^{d,e}$,}
{\large \bf T. Pierzcha\l a$^{f}$,}\\
\vspace{2 mm}
{\large \bf E. Richter-W\c{a}s$^{b,g}$}
{\large \bf Z. W\c{a}s$^{b,h}$} {\large \bf and}
~{\large \bf M. Worek$^{b,i}$ }

\vspace{4 mm} 

{\small
{\em $^a$ CERN, EP/ATR, CH-1211 Geneva 23,
Switzerland.}\\
{\em $^b$ Inst. of Nuclear Physics  PAS, Radzikowskiego 152, 
31-342 Cracow, Poland.}\\ 
{\em $^c$ Faculty of Nuclear Physics and Techniques, AGH, Reymonta 19, 30-059 Cracow, Poland.}\\
 {\em $^d$ Faculty of Mathematics and Physics, University of Ljubljana,\\
Jadranska 19, SI-1000 Ljubljana, Slovenia.}\\
{\em $^e$ Jozef Stefan Institute, Jamova 39, SI-1000 Ljubljana, Slovenia.}\\
{\em $^f$ Inst. of Physics, Univ. of
Silesia, Uniwersytecka 4,  40-007 Katowice, Poland.}\\ 
{\em $^g$Inst. of Physics, Jagellonian University,  Reymonta 
4, 30-059 Cracow, Poland.}\\
{\em $^h$ CERN, Theory Division, CH-1211 Geneva 23, Switzerland.}\\
{\em $^i$ Institute of Nuclear Physics, NCSR `Demokritos', 15310, Athens, Greece.}\\}
\end{center}

\vspace{2mm}
\begin{abstract}
We present the system for versioning two packages: the {\tt TAUOLA}
of $\tau$-lepton decay  and {\tt PHOTOS} for radiative corrections in
decays.  The following features can be chosen in an automatic or
semi-automatic way: (1) format of the  common block {\tt HEPEVT}; 
(2) version of the physics input (for {\tt TAUOLA}): as published, as
initialized by the CLEO collaboration,  as initialized  by the ALEPH
collaboration (it is suggested to use this  version only with the help
of the collaboration advice), new optional parametrizations of
matrix elements in  4$\pi$  decay channels;  
(3) type of application: stand-alone, universal
interface based on the  information stored in the  {\tt HEPEVT}
common block including longitudinal spin effects in the  
elementary $Z/\gamma^{*}\to \tau^+\tau^-$  process, extended version 
of the standard universal interface including full spin effects 
in the  $H/A \to\tau^+ \tau^-$ decay, interface for {\tt KKMC} 
Monte Carlo,  (4) random number generators; (5) compiler options.

\end{abstract}
\vspace{2mm}

\begin{center}
{\it To be submitted to Comput. Phys. Commun.}
\end{center}

\vspace{2 mm}
\footnoterule
\noindent
{\footnotesize
\begin{itemize}
\item[${\dag}$] This work is partly supported by the Polish State
Committee for Scientific Research (KBN) grants Nos. 2P03B00122,
2P03B13025, 5P03B09320, and the European Community's Human Potential
Programme under contracts HPRN-CT-2000-00149 (`Physics at Colliders')
and a Marie Curie Fellowship HPMD-CT-2001-00105 
(`Multi-particle production and higher order correction').

\end{itemize}
}

\end{titlepage}
%%%%%%%%%%%%%%%%%%%%%%%%%%%%%%%%%%%%%%%%%%%%%%%%%%%%%%%%%%%%%%%%%%%%%%%%%%%%%%%%%
%%%%%%%%%%%%%%%%%%%%%%%%%%%%%%%%%%%%%%%%%%%%%%%%%%%%%%%%%%%%%%%%%%%%%%%%%%%%%%%%%
%%%%%%%%%%%%%%%%%%%%%%%%%%%%%%%%%%%%%%%%%%%%%%%%%%%%%%%%%%%%%%%%%%%%%%%%%%%%%%%%%

\noindent{\bf PROGRAM SUMMARY}
\vspace{10pt}

\noindent{\sl Title of the environment:} \- {\tt tauola-photos-F}, release II.

\noindent{\sl Computer:}\-
PC running GNU/Linux operating system

\noindent{\sl Programming languages and tools used:}\-
{\tt CPP}: standard C-language preprocessor, \\
{\tt GNU Make} builder tool,
also {\tt FORTRAN-77} compiler.

\noindent{\sl Size of the package:}\-
 10 MB uncompressed, \\
 2 MB   packed for distribution into {\tt tar.gz} archive.

\noindent{\sl Keywords:}\\
%-----------------------
particle physics, Monte Carlo methods, tau decays, TAUOLA [1], PHOTOS [2]

\noindent{\sl Nature of the physical problem:}\\
The code of Monte Carlo generators often has to be tuned to the needs
of large HEP Collaborations and experiments. Usually, these modifications do not 
introduce important changes in the algorithm, but rather modify the initialization
and form of the hadronic current in $\tau$ decays. The format of the event record 
 ({\tt HEPEVT} common block) used to exchange 
information between building blocks of Monte Carlo systems  often needs
modification. Thus,
there is a need to maintain various, slightly modified versions of the 
same code. The package presented here allows the production of
ready-to-compile versions of {\tt TAUOLA} [1] and {\tt PHOTOS} [2] Monte Carlo 
generators with appropriate demonstration programs.

The new algorithm, universal interface of {\tt TAUOLA} to work with the {\tt HEPEVT} 
common block is also documented here.  Finally  minor technical improvements
of  {\tt TAUOLA} and {\tt PHOTOS} are also listed.

\noindent{\sl Method of solution:}\\
%---------------------------------
The standard {\tt UNIX} tool: the C-language preprocessor is used to produce 
a ready-to-distribute version of {\tt TAUOLA} [1] and {\tt PHOTOS} [2] code.
The final version of {\tt FORTAN77} code is produced from the library of 
`pre-code' that is included in the package.

\noindent{\sl Typical running time:}\\
%-----------------------------------
Depends on the speed of the computer used and the demonstration 
program chosen. Typically a few seconds.

\noindent{\sl Accessibility: } \\
%-----------------------------------
web page: {\tt http://cern.ch/wasm/goodies.html} \\
%or { \tt http://cern.ch/Piotr.Golonka/MC/tauola-photos-F} Piotrek usuwam: 
% chyba ze deklarujesz ze bedziesz
% serwisowal rozne poprawki do formfaktorow TAUOLI itp ... 

\noindent{\sl References: } 
\begin{itemize}
\item `The tau decay library TAUOLA: Version 2.4'\cite{Jadach:1993hs}
\item `PHOTOS: A Universal Monte Carlo for QED radiative corrections. Version 2.0" \cite{Barberio:1994qi}
\end{itemize}

\newpage
%%%%%%%%%%%%%%%%%%%%%%%%%%%%%%%%%%%%%%%%%%%%%%%%%%%%%%%%%%%%%%%%%%%%%%%%%%%%%%%%%%%%%%%%%%%%%
%%%%%%%%%%%%%%%%%%%%%%%%%%%%%%%%%%%%%%%%%%%%%%%%%%%%%%%%%%%%%%%%%%%%%%%%%%%%%%%%%%%%%%%%%%%%%

\tableofcontents
\newpage
%%%%%%%%%%%%%%%%%%%%%%%%%%%%%%%%%%%%%%%%%%%%%%%%%%%%%%%%%%%%%%%%%%%%%%%%%%%%
%%%%%%%%%%%%%%%%%%%%%%%%%%%%%%%%%%%%%%%%%%%%%%%%%%%%%%%%%%%%%%%%%%%%%%%%%%%%
\section{Introduction}
%%%%%%%%%%%%%%%%%%%%%%%%%%%%%%%%%%%%%%%%%%%%%%%%%%%%%%%%%%%%%%%%%%%%%%%%%%%%
%%%%%%%%%%%%%%%%%%%%%%%%%%%%%%%%%%%%%%%%%%%%%%%%%%%%%%%%%%%%%%%%%%%%%%%%%%%%

The {\tt TAUOLA} \cite{Jadach:1990mz,Jezabek:1991qp,Jadach:1993hs} and
{\tt PHOTOS}\cite{Barberio:1990ms,Barberio:1994qi} programs are computing
projects that have a rather long history. Written and maintained by well
defined authors, they nonetheless migrated into a wide range of applications
where they became ingredients of complicated simulation chains. As a
consequence, a large number of different versions are currently in use.
From the algorithmic point of view, they often differ only in a few
small details, but incorporate substantial amount of specific results
from the distinct $\tau$-lepton measurements. 
The difference in the versions of the programs used is often due to
the specific requirements of the interfaces to other packages in the
simulation chain (for instance, the format of the event record
has to be adjusted).

The present utility  for constructing specific versions of {\tt
TAUOLA} and  {\tt PHOTOS} is prepared for software librarians and
advanced users.  The idea was to create a repository which
allows them to include and keep main  options of {\tt TAUOLA} developed for
different purposes. At the same time, the repository can provide the
standard {\tt FORTRAN} files, which can be handled later in  the same
way as the published versions of the packages. The other,
relatively new project, which we will call
{\tt TAUOLA universal interface},  is also discussed 
here.

Our present document is not aimed to be the manual of the  {\tt
PHOTOS} and {\tt TAUOLA} packages.  It is assumed that the  user is
familiar with the programs themselves and their documentation, see Refs.
\cite{Jadach:1990mz,Jezabek:1991qp,Jadach:1993hs,Barberio:1990ms,Barberio:1994qi}. 
The purpose of the paper is to
document how all parts of the code, including different  versions 
and options, are combined into one distribution package. Minor 
improvements in the codes, since the time of publication, are also 
listed here. The paper is an extended and enlarged version of the earlier 
documentation~\cite{Golonka:2000iu}  and  supersedes it. 

\vskip 3 mm

{\bf Motivations for versioning}:

\begin{enumerate}
\item
{\tt PHOTOS}: Diferent versions of the Fortran code are necessary according to the
different  versions of the {\tt HEPEVT} common block being in use in
the HEP libraries  (single/double precisions,  dimension of matrices).
\item
{\tt TAUOLA}: Different versions of Fortran code are motivated by: (A) different
        versions of initialization of physics parameters; (B)
        interfaces with different Monte Carlo generators for
        the production of  $\tau$ lepton(s);  and (C) different versions
        of the {\tt HEPEVT} common block:
        \begin{itemize}
        \item
(A) Different physics initializations:\\  (1) As published in
\cite{Jadach:1993hs}, with improvements from Ref.~\cite{Decker:1996af};
 \\ (2) As initialized by ALEPH collaboration
\cite{aleph} (it is suggested to  use this  version only after asking
the collaboration for advice); \\ (3) As initialized by CLEO
collaboration \cite{cleo} (see printout of this  version for details);
\\ (4) New parallel 4$\pi$ channels of $\tau$ decay based on
Novosibirsk data  \cite{Bondar:2002mw}.
\\ (5) New parallel 4$\pi$ channels of $\tau$ decay based on
Refs. ~\cite{Czyz:2000wh,Decker:1996af}.
\item
(B) Different interfaces with MC generators:\\ (1) Old demo program as
in published version \cite{Jadach:1993hs}; \\ (2) Interface to {\tt
  KKMC} \cite{Jadach:1999vf}; \\ (3) New universal  interface
using {\tt HEPEVT} common block  including spin effects in the
elementary $Z/\gamma^{*} \to \tau^+\tau^-$  process
\cite{Pierzchala:2001gc}.\\ (4) The extended version of
the standard universal  interface including transverse spin
effects in the  $H/A\to \tau^+ \tau^-$ decay 
\cite{Was:2002gv,Bower:2002zx,Desch:2003mw,Desch:2003rw}.
\item
(C) Different versions of the {\tt HEPEVT} common block.\\
\end{itemize}
\item
{\tt TAUOLA} and {\tt PHOTOS}: different versions of random number generators.
\item
{\tt TAUOLA} and {\tt PHOTOS}: {\tt makefile}'s with different compiler flags.
\end{enumerate}

The aim is to provide full backward compatibility at the level of
the Fortran source with the various versions being at present in
use. Standard Unix tools and infrastructure are used in the discussed setup: {\tt cpp} the
C-language preprocessor; its {\tt \#if}, {\tt \#elif} and {\tt \#include}
directives, as well as symbolic links and {\tt cat} command.  It is
expected that the user will employ this setup to create her/his version
of  {\tt TAUOLA} and {\tt PHOTOS} libraries (subdirectories {\tt
tauola/} and {\tt photos/}) and that other subdirectories of the setup will
be erased/stored  separately.

%%%%%%%%%%%%%%%%%%%%%%%%%%%%%%%%%%%%%%%%%%%%%%%%%%%%%%%%%%%%%%%%%%%%%%%%%%%%
%%%%%%%%%%%%%%%%%%%%%%%%%%%%%%%%%%%%%%%%%%%%%%%%%%%%%%%%%%%%%%%%%%%%%%%%%%%%
\section{Organization of the directory tree}
%%%%%%%%%%%%%%%%%%%%%%%%%%%%%%%%%%%%%%%%%%%%%%%%%%%%%%%%%%%%%%%%%%%%%%%%%%%%
%%%%%%%%%%%%%%%%%%%%%%%%%%%%%%%%%%%%%%%%%%%%%%%%%%%%%%%%%%%%%%%%%%%%%%%%%%%%

Once unpacked, the main directory  { \tt TAUOLA} is created.

\noindent
It contains {\tt README} file and:

\vskip 2mm

\centerline{ \bf subdirectories including Fortran precode}

\vskip 2mm

\begin{enumerate}

\item
{\tt photos-F/}: Main directory containing PHOTOS precode with
options.
\item
{\tt tauola-F/}: Main directory containing TAUOLA precode with
options.
\item
{\tt tauola-factory/}: Directory containing the  {\tt glue} program,
which prepares the nonstandard directory {\tt tauola-F/}.
\item
{\tt demo-factory/}: Directory for updating the input in demo files. For
specialized  use only, see Section \ref{setting}.
\item
{\tt randg/}: Directory containing random number generators, which are
kept  separately from the rest of {\tt TAUOLA} and {\tt PHOTOS} source
code.  This should facilitate the replacement with the versions of random
number  generators  favoured by the user. Random number generators are
kept in Fortran subroutines placed in files {\tt photos-random.h} and
{\tt tauola-random.h}.
\item
{\tt include/}: Directory containing the {\tt  HEPEVT-xxx.h} files
for  different versions of the {\tt HEPEVT} common block.  The symbolic
link {\tt HEPEVT.h} to the one actually used will be placed in this
directory at the time of initialization. 
The {\tt  phyfix.xxx} files with versions of Fortran subroutine
{\tt PHYFIX} used only by {\tt TAUOLA universal interface} are also stored there. The
version to be used needs to be copied into  {\tt  phyfix.h};
 for more details, see {\tt  README-phyfix}.
\end{enumerate}

\vskip 2mm
\centerline{ \bf subdirectories necessary to run demo and}
\centerline{ \bf to install packages on different platforms}

\vskip 2mm

\begin{enumerate}
\item
{\tt glibk/}: Directory containing histogramming package, used by demos
only.
\item
{\tt jetset/}: Directory containing {\tt JETSET} MC package, used by
demos  only.
\item
{\tt jetset2/}: Directory containing {\tt PYTHIA} MC package, used by
demos  only.
\item
{\tt eli/}: Directory containing files necessary to construct demo to be run with
{\tt PYTHIA} MC package.
\item
{\tt platform/}: Directory containing system-dependent versions of
                    {\tt make.inc} files for supported  platforms.
\item
{\tt make.inc}: Symbolic link to the chosen {\tt make-xxxx.inc}
located in subdirectory {\tt platform/}. The {\tt make-xxxx.inc} files
define machine-dependent flags for compilers etc. to be used by
all {\tt makefiles}.
\end{enumerate}

\vskip 2mm \centerline{\bf The following directories are created}
\centerline{\bf once the actions of the setup are completed:}
\vskip 2mm

\begin{enumerate}
  \item
  {\tt photos/}: Standard directory with Fortran code of {\tt PHOTOS}
  library and its demo.
  \item
  {\tt tauola/}: Standard directory with Fortran code of {\tt TAUOLA}
                    library,  its demos and example outputs.
  \begin{enumerate}
    \item
    {\tt tauola/demo-standalone}: Demo program for {\tt TAUOLA}
executed in a  standalone mode.
    \item
    {\tt tauola/demo-jetset}: Demo program for {\tt  TAUOLA} executed
    with  {\tt universal interface} and physics event generators using
    the {\tt HEPEVT} common block. In this demo  {\tt HEPEVT} is
    filled by {\tt JETSET74} \cite{Sjostrand:1987hx} Monte Carlo
    event generator.
    \item
    {\tt tauola/demo-pythia}: Optionally, a demo program for {\tt  TAUOLA} executed
    with {\tt  universal interface} and {\tt HEPEVT} common block
    filled by {\tt PYTHIA} \cite{Sjostrand:1987hx} Monte Carlo event
    generator can be created as well.

   \item
   {\tt tauola/demo-KK-face}: Interface to KK Monte Carlo
   \cite{Jadach:1999vf}.
  \end{enumerate}
\end{enumerate}

%%%%%%%%%%%%%%%%%%%%%%%%%%%%%%%%%%%%%%%%%%%%%%%%%%%%%%%%%%%%%%%%%%%%%%%%%%%%
%%%%%%%%%%%%%%%%%%%%%%%%%%%%%%%%%%%%%%%%%%%%%%%%%%%%%%%%%%%%%%%%%%%%%%%%%%%%
\subsection{Options for {\tt PHOTOS} Monte Carlo}
%%%%%%%%%%%%%%%%%%%%%%%%%%%%%%%%%%%%%%%%%%%%%%%%%%%%%%%%%%%%%%%%%%%%%%%%%%%%
%%%%%%%%%%%%%%%%%%%%%%%%%%%%%%%%%%%%%%%%%%%%%%%%%%%%%%%%%%%%%%%%%%%%%%%%%%%%
\label{OpPHOTOS}
The different options of {\tt PHOTOS} that can be created, correspond solely to
the different versions of the {\tt HEPEVT} common block.  The possible
options are:
\begin{enumerate}
\item
{\tt KK-all}     -- for KK Monte Carlo, older versions
\item
{\tt 2kD-all}    -- arrays dimension  2000, double precision
\item
{\tt 4kD-all}    -- arrays dimension  4000, double precision
\item
{\tt 2kR-all}    -- arrays dimension  2000, single precision
\item
{\tt 10kD-all}   -- arrays dimension 10000, double precision
\end{enumerate}

The action of preparing the required version of the library is performed
with the help of  {\tt cpp } preprocessor. It creates a file {\tt photos.f} from
precode stored in {\tt photos.F}. Once this is done,  the symbolic link to  the
required version of the {\tt HEPEVT} common block is defined.  This
link is used in the construction of the {\tt tauola} library, see next
section.

%%%%%%%%%%%%%%%%%%%%%%%%%%%%%%%%%%%%%%%%%%%%%%%%%%%%%%%%%%%%%%%%%%%%%%%%%%%%
%%%%%%%%%%%%%%%%%%%%%%%%%%%%%%%%%%%%%%%%%%%%%%%%%%%%%%%%%%%%%%%%%%%%%%%%%%%%
\subsection{Options for {\tt TAUOLA} Monte Carlo}
%%%%%%%%%%%%%%%%%%%%%%%%%%%%%%%%%%%%%%%%%%%%%%%%%%%%%%%%%%%%%%%%%%%%%%%%%%%%
%%%%%%%%%%%%%%%%%%%%%%%%%%%%%%%%%%%%%%%%%%%%%%%%%%%%%%%%%%%%%%%%%%%%%%%%%%%%

Basic options for physics initializations  are: {\tt cpc}; {\tt cleo};
{\tt aleph}. As results of the action performed by the package:
\begin{enumerate}
\item
The {\tt tauola/} subdirectory is erased;
\item
 The directory structure of  {\tt tauola/} is rebuilt;
\begin{itemize}
\item
 the {\tt tauola/} directory is filled with the Fortran code, libraries
and  makefiles;
\item
 {\tt tauola/demo-xx} are filled with the Fortran code of demos;
\end{itemize}
\end{enumerate}

The three possible versions of created {\tt tauola.f} correspond to
formfactors and branching ratios defined respectively as in:  ({\tt cpc})
published version of TAUOLA\cite{Jadach:1993hs,Decker:1996af}; ({\tt aleph}) as adopted by ALEPH
collaboration\cite{aleph},  ({\tt cleo}) as adopted by CLEO collaboration\cite{cleo}.\\

{\bf Remarks:}\\

   $\bullet$ The {\tt makefile} files are prepared to run {\tt TAUOLA}
   within the environment of the distribution {\tt TAUOLA} directory;
   however the templates for makefiles are compatible with those of
   the {\tt KK} Monte Carlo. Thus if the {\tt tauola} directory  is copied
   into its respective place in the {\tt KKMC} distribution tree, and
   {\tt make makflag} of {\tt KK/ffbench/} is executed, it overwrites
   the {\tt  makefile}  file in {\tt tauola/}. The new ones are  produced
   from {\tt makefile.templ} and  match the {\tt KKMC} structure.

   $\bullet$ Additional parametrizations for formfactors, which can
   be useful in some applications, are stored in the directory {\tt
   TAUOLA/tauola-F/suppl}. They are not ready to use and some
   cross-checks are mandatory. At present, the code used in
   Refs. \cite{Abbiendi:1999cq} and \cite{Abreu:1998cn}  is stored
   there.

%%%%%%%%%%%%%%%%%%%%%%%%%%%%%%%%%%%%%%%%%%%%%%%%%%%%%%%%%%%%%%%%%%%%%%%%%%%%
%%%%%%%%%%%%%%%%%%%%%%%%%%%%%%%%%%%%%%%%%%%%%%%%%%%%%%%%%%%%%%%%%%%%%%%%%%%%
\subsection{How to change the setting of {\tt TAUOLA} input parameters}
%%%%%%%%%%%%%%%%%%%%%%%%%%%%%%%%%%%%%%%%%%%%%%%%%%%%%%%%%%%%%%%%%%%%%%%%%%%%
%%%%%%%%%%%%%%%%%%%%%%%%%%%%%%%%%%%%%%%%%%%%%%%%%%%%%%%%%%%%%%%%%%%%%%%%%%%%
\label{setting}

It is often necessary to change some of the {\tt TAUOLA}  input
parameters, such as branching ratios, mass of the $\tau$-lepton,
etc. It is convenient to have it done  once for  all applications,
i.e. {\tt demo-KK-face}, {\tt demo-jetset} and {\tt  demo-standalone}.
The purpose of the {\tt demo-factory} directory is exactly that.  Here
one can create the {\tt .F} files for the interfaces, by the set of
`Paste' commands embodied in the script {\tt klej},  out of the blocks
of Fortran code.  More precisely the following files can be
re-created:
\begin{itemize}
\item
For {\tt demo-KK-face}:                 {\tt ./prod/Tauface.F}
\item
For {\tt demo-jetset}:        {\tt ./prod/tauola$\_$photos$\_$ini.F}
\item
For {\tt demo-standalone}:               {\tt ./prod/taumain.F}
\end{itemize}

For details of the intialization routines, which are semi-identical in
the  three cases,  see Refs.
\cite{Jadach:1990mz,Jezabek:1991qp,Jadach:1993hs}.  This requires
special care from the physics point of view. In many cases, the input
parameters  are inter-related with the actual choice  of form
factors. The changes should thus be performed consistently.\\

{\bf How to proceed:}\\
\begin{enumerate}
\item
   Some of the routines in directory {\tt ./source}  have to be
   updated by hand first. They are stored in individual files.  The
   ones that usually should not be modified are write-protected.
\item
   Later execution of the script {\tt klej} will create the following
   files from the pieces stored in the directory {\tt ./source} simply by
   pasting them  together:
\begin{itemize}
 \item
 {\tt  ./prod/Tauface.F},
\item
 {\tt  ./prod/tauola$\_$photos$\_$ini.F},
\item
 {\tt ./prod/taumain.F}.
\end{itemize}
  Automatic check (using {\tt diff}) with the archive versions stored in
  the directory {\tt ./back}  will also be executed.
\item
  Finally the following commands copy the files into the appropriate
  places:
  \begin{enumerate}
    \item
    {\tt cp  prod/Tauface.F ../tauola-F/tauface-KK-F/Tauface.F}
    \item
    {\tt cp  prod/tauola$\_$photos$\_$ini.F
../tauola-F/jetset-F/tauola$\_$photos$\_$ini.F}
    \item
    {\tt cp  prod/taumain.F ../tauola-F/standalone-F/taumain.F}
  \end{enumerate}
\end{enumerate}

%%%%%%%%%%%%%%%%%%%%%%%%%%%%%%%%%%%%%%%%%%%%%%%%%%%%%%%%%%%%%%%%%%%%%%%%%%%%
%%%%%%%%%%%%%%%%%%%%%%%%%%%%%%%%%%%%%%%%%%%%%%%%%%%%%%%%%%%%%%%%%%%%%%%%%%%%
\subsection{Random number generators}
%%%%%%%%%%%%%%%%%%%%%%%%%%%%%%%%%%%%%%%%%%%%%%%%%%%%%%%%%%%%%%%%%%%%%%%%%%%%
%%%%%%%%%%%%%%%%%%%%%%%%%%%%%%%%%%%%%%%%%%%%%%%%%%%%%%%%%%%%%%%%%%%%%%%%%%%%
\label{randnumgen}

 \begin{itemize}
 \item
  {\tt PHOTOS} and {\tt TAUOLA} have their own copies of the random
number  generators.  They are contained in the {\tt include} files placed in
the directory {\tt randg}.
 \item
  The user who wants to implement her/his own generators, e.g.
 compatible with the ones used by the collaboration, should replace
 the files:
 \begin{itemize}
 \item 
    {\tt ./photos-random.h},
 \item 
    {\tt ./tauola-random.h}
 \end{itemize}
  by those including the appropriate wrappers of his/her own random
  generators (or empty files if the generators of the same name reside
  elsewhere).
\end{itemize} 

%%%%%%%%%%%%%%%%%%%%%%%%%%%%%%%%%%%%%%%%%%%%%%%%%%%%%%%%%%%%%%%%%%%%%%%%%%%%
%%%%%%%%%%%%%%%%%%%%%%%%%%%%%%%%%%%%%%%%%%%%%%%%%%%%%%%%%%%%%%%%%%%%%%%%%%%%
\subsection{Compiler flags, etc.}
%%%%%%%%%%%%%%%%%%%%%%%%%%%%%%%%%%%%%%%%%%%%%%%%%%%%%%%%%%%%%%%%%%%%%%%%%%%%
%%%%%%%%%%%%%%%%%%%%%%%%%%%%%%%%%%%%%%%%%%%%%%%%%%%%%%%%%%%%%%%%%%%%%%%%%%%%

Platform-dependent parts of the {\tt makefiles} are stored in
the directory {\tt platform/}.  At present options are available for various
distributions of the Linux platform only,  
but it is rather straightforward to extend them to the new ones.

%%%%%%%%%%%%%%%%%%%%%%%%%%%%%%%%%%%%%%%%%%%%%%%%%%%%%%%%%%%%%%%%%%%%%%%%%%%%
%%%%%%%%%%%%%%%%%%%%%%%%%%%%%%%%%%%%%%%%%%%%%%%%%%%%%%%%%%%%%%%%%%%%%%%%%%%%
\section{Universal interface with {\tt HEPEVT} common block}
%%%%%%%%%%%%%%%%%%%%%%%%%%%%%%%%%%%%%%%%%%%%%%%%%%%%%%%%%%%%%%%%%%%%%%%%%%%%
In the present section, we shall document for the first time, 
how to install and use the {\tt TAUOLA universal interface }
 with `any' production generator, to  include
 spin effects in e.g. the $Z/\gamma^{*} \to \tau^+\tau^-$
 process \cite{Pierzchala:2001gc,Was:2002gv}.  
 The approximate spin
 correlations are  calculated from the information stored in the {\tt
 HEPEVT}  common block \cite{Caso:1998tx} filled by `any' $\tau$
 production program.  As a demonstration example it is interfaced with
 the {\tt JETSET} generator; however, it  works in the same manner
 with {\tt PYTHIA, HERWIG} and should work  with the {\tt ISAJET} generator as well.  
 In fact, such
 an  interface can be considered as a separate software project, to
 some degree independent from both the specific problem of $\tau$
 production and its decay.  The aim of this interface  is not to
 replace the matrix element calculations, but rather to provide a
 method  of calculating/estimating spin effects in cases when these
 would not be taken into account at all.

%%%%%%%%%%%%%%%%%%%%%%%%%%%%%%%%%%%%%%%%%%%%%%%%%%%%%%%%%%%%%%%%%%%%%%%%%%%%
\subsection{{\tt HEPEVT} common block and {\tt HEPEVT} event record}
%%%%%%%%%%%%%%%%%%%%%%%%%%%%%%%%%%%%%%%%%%%%%%%%%%%%%%%%%%%%%%%%%%%%%%%%%%%%

%%%%%%%%%%%%%%%%%%%%%%%%%%%%%%%%%%%%%%%%%%%%%%%%%%%%%%%%%%%%%%%%%%%%%%%%%%%%
The question of adapting the universal interface%
\footnote{The question is even more serious in the case 
of {\tt PHOTOS}.} 
to different versions of  an {\tt HEPEVT} event record goes
beyond the technicalities discussed in Section \ref{OpPHOTOS}. In fact it is quite
involved, as it depends on specific needs due to the way the 
{\tt HEPEVT event record}
is actually coded into the {\tt HEPEVT common block}. It varies from application to 
application according to the physics requirement. 
During the {\tt MC4LHC} Workshop~\cite{MC4LHC} these issues were discussed and it was found, thanks
to interaction with the users,  that the {\tt TAUOLA universal interface} works
with all 3 basic options for {\tt HEPEVT} filled by {\tt PYTHIA} 6.2 
({\tt MSTP(128)=0,1,2}), as well 
as with {\tt HERWIG} (6.150) \cite{HERWIG}. 
In our example we keep only the working interface with 
{\tt PYTHIA }version
5.7, the working example with other installations can be found 
e.g. in~\cite{AcerMC}. 

The discussion of the related problems go
beyond this document; we refer the reader to Section \ref{sekcja8} for more details.
We are also convinced that the problems, because they originate from physics, will 
appear in event records in other programming languages such as C++.

%%%%%%%%%%%%%%%%%%%%%%%%%%%%%%%%%%%%%%%%%%%%%%%%%%%%%%%%%%%%%%%%%%%%%%%%%%%%
\subsection{Longitudinal spin effects}
%%%%%%%%%%%%%%%%%%%%%%%%%%%%%%%%%%%%%%%%%%%%%%%%%%%%%%%%%%%%%%%%%%%%%%%%%%%%
%%%%%%%%%%%%%%%%%%%%%%%%%%%%%%%%%%%%%%%%%%%%%%%%%%%%%%%%%%%%%%%%%%%%%%%%%%%%

Let us start with the simpler case, when 
only longitudinal spin degrees are included. Then 
the interface acts in the following way:
\begin{itemize}
\item  The user has to verify if 
the $\tau$ lepton is forced to be stable in the package
 performing generation  of the $\tau$ production.
\item
The contents of the {\tt HEPEVT} common block is searched  for all
$\tau$ leptons and $\tau$ neutrinos.
\item
It is checked if there are $\tau$ flavour pairs (two $\tau$ leptons or
a $\tau$ lepton and a $\tau$ neutrino) originating from the same mother(s).
\item
The decays of the $\tau$ flavour pairs are performed with the
subroutine {\tt  TAUOLA}.  Longitudinal spin correlations are
generated  in the case of the $\tau$ produced from the decay of: $W \to
\tau \nu$, $Z/\gamma^{*} \to \tau \tau$, 
and the charged Higgs boson $H^{\pm}\to  \tau \nu$
(for the neutral Higgs boson the full spin correlations are used, see 
Section~\ref{transv}).
Parallel or antiparallel spin configurations are generated, before
calling on the $\tau$ decay, and then the decays of 100\% polarized
$\tau$'s are executed.
\item
In the case of the Higgs boson (for the spin correlations to be
generated) the identifier of the $\tau$ mother must be that of the
Higgs boson.  The particle code convention as that used by the
{\tt PYTHIA 5.7}\cite{Sjostrand:1994yb}  Monte Carlo is adopted, but it can be
changed by the user, as explained later in this section.
\item
In the case of the $W$ and $Z/\gamma^{*}$, this is not necessary, because the
interface assumes these production mechanisms as  defaults, even if the  
$W$ or $Z/\gamma^{*}$ bosons are not 
explicitly present in the event record.  For example 
if from the same mother as that of the $\tau$, a $\nu_\tau$ is also  produced,
the $W$ (of the momentum equal to the sum of $\tau$ and $\nu_\tau$) 
will be taken as the true mother of the  $\tau$.  Similarly, if
another $\tau$ with opposite charge is produced from the same mother, the
$Z/\gamma^{*}$ is assumed to be the mother of the  $\tau$ pair.
\item
The spin effects in the states of $\tau^+$, $\tau^-$ produced from
explicit or  implicit $Z/\gamma^{*}$ states are upgraded up to nearly
{\tt KORALZ} standards \cite{Jadach:1994yv,Jadach:1999tr}.  
If bremsstrahlung protons are present  in the event record (as sisters of $Z$),
then they have to be mergerd either with beams 
or with $\tau$'s, accordingly to approximation explained in 
\cite{Pierzchala:2001gc}. 
Photon radiation in the decay is performed with {\tt PHOTOS} package
\cite{Barberio:1990ms,Barberio:1994qi}.
\end{itemize}
Let us note that the calculation of the $\tau$ polarization created
from the $Z$ and/or virtual $\gamma^{*}$ (as a function of the
direction) represents  a rather non-trivial extension if high precision 
is required.  Generally, the
dedicated study of  the production matrix elements of the host
generator is necessary in every  individual case. The method applied here
is based on approximations \cite{Pierzchala:2001gc,Was:2002gv}.

%%%%%%%%%%%%%%%%%%%%%%%%%%%%%%%%%%%%%%%%%%%%%%%%%%%%%%%%%%%%%%%%%%%%%%%%%%%%
\subsection{Organization of the interface}
%%%%%%%%%%%%%%%%%%%%%%%%%%%%%%%%%%%%%%%%%%%%%%%%%%%%%%%%%%%%%%%%%%%%%%%%%%%%
%%%%%%%%%%%%%%%%%%%%%%%%%%%%%%%%%%%%%%%%%%%%%%%%%%%%%%%%%%%%%%%%%%%%%%%%%%%%

The {\tt TAUOLA} interface is organized in a modular form to be  used
conveniently in `any environment'.
\vskip 0.4 cm \centerline{\bf Initialization}
\vskip 0.4 cm

Initialization is performed with the  {\tt CALL TAUOLA(MODE,KEYSPIN)},
 {\tt MODE=-1}.  All necessary input is directly coded in the subroutine
 {\tt TAUOLA} placed  in a file {\tt tauface-jetset.f}.

The following input parameters are set at this call. We omit those,
which are standard input for {\tt TAUOLA} as defined in its
documentation.  They are hard-coded in the subroutine.

\def\sstrut{$\strut\atop\strut$}
\vskip 0.1 cm
%========================================
\vbox{
$$\halign{ \vrule # & \hskip5pt  \sstrut  {\tt #} \hfil \vrule &
\hskip5pt  \vtop{\hsize=11.8cm {\noindent \strut # \strut}} & #
\vrule\cr \noalign{\hrule} 
& Parameter     & Meaning                                                          &\cr
\noalign{\hrule} 
&POL            & Internal switch for spin effects in $\tau$
                  decays and only for the longitudinal case. Normally the user should set {\tt POL=1.0}, 
                  and when {\tt
                  POL=0.0} spin polarization effects are absent.                           &\cr
&KFHIGGS(3)     &({\tt KF}=25, 35, 36) Flavour code for $h$, $H$ and $A$                  &\cr 
&KFHIGCH        &({\tt KF}=37) Flavour code for $H^{+}$                                   &\cr 
& KFZ0          &({\tt KF}=23) Flavour code for $Z^{0}$                                   &\cr 
& KFGAM         &({\tt KF}=22) Flavour code for $\gamma$                                  &\cr 
& KFTAU         &({\tt KF}=15) Flavour code for $\tau^{-}$                                &\cr 
& KFNUE         &({\tt KF}=16) Flavour code for $\nu_{\tau}$                              &\cr 
& xmtau         & $\tau$-lepton mass, used only for {\tt KFHIGGS(3)}, see Section \ref{transv} &\cr
& xmh           & Higgs boson mass, used only for {\tt KFHIGGS(3)}                        &\cr
& psi           & Mixing scalar--pseudoscalar angle, this angle is used only for 
                  Higgs boson with flavour code equal {\tt KFHIGGS(3)}                    &\cr
\noalign{\hrule} }$$}
 
%\textcolor{magenta}{\centerline{ For ZbW:   /home/wasm/y2002/tauola-code/TAUOLA/tauola/demo-pythia}}
 
\vskip 0.4 cm \centerline{\bf Event generation}
\vskip 0.4 cm

For every event generated by the production generator,  
all $\tau$ leptons will decay with the single {\tt CALL
TAUOLA(0,KEYSPIN)}  ({\tt KEYSPIN=1/0} denotes spin effects switched
on/off). In its execution, all $\tau$ leptons will be first localized, their
positions stored in internal common block {\tt TAUPOS}, and the
information necessary for calculating  the $\tau$ spin state will be
read from {\tt HEPEVT} common block.  The spin state for the
given  $\tau$  (or  $\tau$ pair)  will be generated later, and finally the decay
of the polarized $\tau$ will be   performed with the standard  {\tt
TAUOLA} action. In  particular, the decay products of $\tau$ will be
boosted to the laboratory  frame  and added to the complete event
configuration stored  in the {\tt HEPEVT} common  block.

\vskip 0.4 cm \centerline{\bf Calculation of the $\tau$ spin state}
\vskip 0.4 cm

Once {\tt CALL TAUOLA(0,KEYSPIN)} is executed and $\tau$ leptons
are found,  the spin states have to be calculated.

First, we look  in {\tt HEPEVT} for the position of  $\tau$'s mothers
and store  them in matrix {\tt IMOTHER(20)}. Each mother giving
$\tau$ lepton(s) is stored  only once, independently of the number of
produced  $\tau$'s.  Later, for every {\tt IMOTHER(i)} we execute the
following steps:

\begin{enumerate}
\item
 The daughters, which are either $\tau$ leptons or $\nu_\tau$, are
 searched for.
\item
 The daughters are combined in pairs, and the case of more than 1 pair is not
 expected  to be important; ad hoc pairing is then performed.
\item
 The two main cases are thus ($\tau $  $\tau $)  or ($\tau$ $\nu_\tau $).
\item
 The default choices in these two main cases are, respectively, 
$Z/\gamma^{*}$ or $W$, unless
the identifier of {\tt IMOTHER(i)} is different and explicitly defined,
e.g. as a neutral (or charged) Higgs boson.
\item 
Calculation of the spin parameters is independent from kinematic variables and
straightforward in all cases except  $Z/\gamma^{*}$ (see \cite{Pierzchala:2001gc}).
\item
For $Z/\gamma^{*}$, the polarization function $P_{Z}$ 
\cite{Pierzchala:2001gc} is calculated with the help of the
function  {\tt PLZAPX(HOPE,IM0,NP1,NP2)}. The {\tt HOPE} is the
logical parameter  defined in subroutine {\tt TAUOLA} placed in the file
{\tt tauface-jetset.f}.  It tells  whether spin effects can
be calculated or not. If available
information is incomplete,  {\tt HOPE} is set to {\tt  .false.}. Then  {\tt PLZAPX(HOPE,IM0,NP1,NP2)}
returns $0.5$. The  {\tt IM0} denotes the position  of the  $\tau$ mother
in {\tt HEPEVT} common block, the {\tt NP1} denotes the position of  $\tau^{+}$ and
the  {\tt NP2} denotes  the position of  $\tau^-$.
\item
  To calculate the reduced $2 \to 2 $ body kinematical variables $s$ and
$\cos\theta$, the  subroutine {\tt ANGULU(PD1,PD2,Q1,Q2,COSTHE)} is used.
Four-momenta of the incoming effective beams and outgoing  $\tau^+$ and
$\tau^-$ are denoted by $\tt PD1, PD2, Q1, Q2$, respectively.
The variables  $s$ and $\cos\theta$ are then used in the function $P_{Z}$.
\end{enumerate}

\vskip 0.4 cm \centerline{\bf Run summary}
\vskip 0.4 cm

  After the series of events is generated the optional {\tt CALL
  TAUOLA(1,KEYSPIN)} can be executed.  The  information on the whole
  sample, such as the number of generated $\tau$ decays, branching
  ratios calculated from matrix elements, etc.  will be printed.

\vskip 0.4 cm \centerline{\bf Demonstration program}
\vskip 0.4 cm

Our main program  {\tt demo.f} is stored in the subdirectory {\tt
demo-jetset}. It  reads in the file {\tt init.dat}, which includes some
input parameters for the  particular run, such as the number of events to
be generated (by {\tt JETSET/PYTHIA}), or the  type of the interaction
it should use to produce $\tau$'s. We address the reader
directly to the code for more details. It is self-explanatory.
The printouts at the interface initialization has the form presented in
Fig. \ref{tauola-init-printout}, while that at the end of the execution 
has the form presented in Fig. \ref{tauola-end-printout}.

\begin{figure}
{\small
\begin{alltt}
 ***************************************************************************
 *                         *****TAUOLA UNIVERSAL INTERFACE: ******         *
 *                         *****VERSION 1.10, November 2003 ******         *
 *                         **AUTHORS: P. Golonka, B. Kersevan, ***         *
 *                         **T. Pierzchala, E. Richter-Was, ******         *
 *                         ****** Z. Was, M. Worek ***************         *
 *                         **USEFUL DISCUSSIONS, IN PARTICULAR ***         *
 *                         *WITH C. Biscarat and S. Slabospitzky**         *
 *                         ****** are warmly acknowledged ********         *
 *                                                                         *
 *                         ********** INITIALIZATION  ************         *
 *             1.00000     tau polarization switch must be 1 or 0          *
 *             1.57080     Higs scalar/pseudo mix CERN-TH/2003-166         *
 ***************************************************************************
\end{alltt} 
}
\caption{ Printout of {\tt TAUOLA universal interface} initialization}
\label{tauola-init-printout}
\end{figure}

\noindent
\begin{figure}
{\small
\begin{alltt}
 ***************************************************************************
 *                         *****TAUOLA UNIVERSAL INTERFACE: ******         *
 *                         *****VERSION 1.10, November 2003 ******         *
 *                         **AUTHORS: P. Golonka, B. Kersevan, ***         *
 *                         **T. Pierzchala, E. Richter-Was, ******         *
 *                         ****** Z. Was, M. Worek ***************         *
 *                         **USEFUL DISCUSSIONS, IN PARTICULAR ***         *
 *                         *WITH C. Biscarat and S. Slabospitzky**         *
 *                         ****** are warmly acknowledged ********         *
 *                         ****** END OF MODULE OPERATION ********         *
 ***************************************************************************
\end{alltt}
}
\caption{ Printout of {\tt TAUOLA universal interface} end-of-run messages}
\label{tauola-end-printout}
\end{figure}

%%%%%%%%%%%%%%%%%%%%%%%%%%%%%%%%%%%%%%%%%%%%%%%%%%%%%%%%%%%%%%%%%%%%%%%%%%%%
%%%%%%%%%%%%%%%%%%%%%%%%%%%%%%%%%%%%%%%%%%%%%%%%%%%%%%%%%%%%%%%%%%%%%%%%%%%%
\subsection{Transverse spin effects}
%%%%%%%%%%%%%%%%%%%%%%%%%%%%%%%%%%%%%%%%%%%%%%%%%%%%%%%%%%%%%%%%%%%%%%%%%%%%
%%%%%%%%%%%%%%%%%%%%%%%%%%%%%%%%%%%%%%%%%%%%%%%%%%%%%%%%%%%%%%%%%%%%%%%%%%%%
\label{transv}

The extended version of the standard {\tt universal interface}, of the
{\tt TAUOLA} $\tau$-lepton decay library,  includes the complete spin
effects for $\tau$ leptons originating  from the spin zero particle, i.e. in
$H/A \to \tau^+ \tau^-$ decay 
\cite{Was:2002gv,Bower:2002zx,Desch:2003mw}.   As usual, the
interface is expected to work with any Monte Carlo generator providing
production (and subsequent decay  into pair of $\tau$ leptons) of the
Higgs boson.   Since the Higgs boson spin is zero, the spin
correlations of its decay products do not depend at all  on the
mechanism of the Higgs boson production.  Technical difficulties
related to the choice of $\tau^+$ and $\tau^-$ spin quantization
frames, present in the case   of  $e^+e^- \to  Z/\gamma^{*} \to \tau^+
\tau^-$ \cite{Jadach:1998wp,Jadach:1984ac} (initial state bremsstrahlung effects
included or not), are not present. The analytical form of the density
matrix is simple.  To calculate the density matrix for the pair of
$\tau$-leptons  it is thus enough to: know  their four-momenta, know
that they indeed originate from the Higgs boson, and assume the type
of  the Yukawa interaction. Such information is stored  in the
{\tt HEPEVT} common block \cite{Caso:1998tx}
used by practically all Monte Carlo generators for Higgs boson
production.  
%In  Refs. \cite{Pierzchala:2001gc}, the
%algorithm was developed where all $ \tau$ leptons found in the {\tt
%HEPEVT} common block can be decayed  with the help of the {\tt TAUOLA}
%library \cite{Jadach:1990mz,Jezabek:1991qp,Jadach:1993hs} and the $
%\tau$ decay products are appended to the {\tt HEPEVT} as well.  The
%kinematical information on the momenta of all particles forming an
%event was used to calculate, in some approximation, the longitudinal
%spin state of the $ \tau$.  For our purpose that solution had to be
%extended, to incorporate the full density matrix of the $\tau^+
%\tau^-$ pair, in the case  when it is originating  from the Higgs
%boson decay. 
\\

 The interface acts in the following way and according to principles 
described in \cite{Jadach:1990mz}:
\begin{itemize}
\item
   The algorithm is organized in two steps. In the first
   step, the $\tau$ lepton pair   is generated and the $\tau$ leptons
   decay in their respective    rest frames, as if there    were no spin
   effects at all. In the second step, the spin weight is calculated
   and rejection is performed. If the event is rejected, only    the
   generation of the $\tau$ lepton decays is repeated.
\item
   The spin weight for Higgs boson decay into $\tau^{+}\tau^{-}$ is
   given by the following formula
   \begin{equation}  
     wt= {1 \over 4} \Bigl( 1 +\sum^{3}_{i=1} 
	\sum^{3}_{j=1} R_{ij} h_{1}^i h_{2}^j \Bigr) .
   \end{equation}  
\item
   The components of the density matrix $R_{ij}$ are equal 
   $R_{33} =-1$, $R_{11} =\pm 1$, $R_{22}=\pm 1$, respectively,
   for pure scalar and
   pseudoscalar Higgs boson, and all other components are   zero. The
   $ h^i$ and $h^j$ are the polarimetric vectors of the $\tau^{\pm}$
   leptons (they are calculated by {\tt TAUOLA}).
\item
   In general, the Yukawa coupling can be written as  $\bar{\tau}(a+ib\gamma_{5})\tau$.
   With our choice of quantization frames, the following non-zero components of $R_{ij}$ are obtained
   (see Ref.~\cite{Desch:2003rw}):
   \begin{equation}
    R_{33}=-1,~~~~~
    R_{11}=R_{22}=\frac{a^{2}\beta^{2}-b^{2}}{a^{2}\beta^{2}+b^{2}},~~~~~~~
    R_{12}=-R_{21}=\frac{2ab\beta}{a^{2}\beta^{2}+b^{2}},
   \end{equation}
   where $\beta=\sqrt{1-\frac{4m^{2}_{\tau}}{m^{2}_{H}}}$. With Yukawa coupling
   written in the form  $\bar{\tau}N(\cos\phi+i\sin\phi\gamma_{5})\tau$ and with  the limit $\beta \to 1$,
   the expressions reduce to the
   components of the rotation  matrix  by an angle $-2\phi$: 
   \begin{equation}
    R_{11}=R_{22}=\cos2\phi, ~~~~~~~ R_{12}=-R_{21}=\sin2\phi.
   \end{equation}
\end{itemize}
\vspace{4 mm}

 The following changes had to be introduced to the
algorithm explained in Ref.~\cite{Pierzchala:2001gc}:

\begin{itemize}
\item
   The quantization frames for the spin states of  $ \tau^+$ and $\tau^-$
   need to be  properly oriented with respect to each other.  In our
   solution  they are simply connected by the boost along the $\tau$ lepton
   momenta as defined in the Higgs boson rest frame. At the technical
   level this is enforced by the  {\tt TRALOR} routine
   \cite{Jadach:1990mz}, which defines the relation  of the $\tau^\pm$ spin
   quantization frames with the laboratory frame. As an intermediate step
   this routine uses the Higgs boson rest frame.
\item
   The density matrix for the pure scalar/pseudoscalar case was taken from Ref. \cite{Kramer:1994jn},
   and in the most general case in  Ref. \cite{Grzadkowski:1995rx}.
   We have adapted it to  the quantization  frames as specified previously.
   Two cases are implemented:  pure scalar Higgs boson and 
   Higgs boson with mixed scalar--pseudoscalar coupling.
\item
    Generation of $\tau^+$ and $\tau^-$ decays is implemented in
    the subroutine {\tt SPINHIGGS} placed  in the file {\tt
    tauface-jetset.f} and  performed following the method explained and used
    in Refs.\cite{Jadach:1990mz,Jadach:1984ac}.
\item  
    The function  {\tt wthiggs(IFPSEUDO,HH1,HH2)}   
    is used to calculate the  spin weight.
    The  {\tt HH1} and {\tt
    HH2} are the $\tau^{+}$ and $\tau^{-}$  polarimetric vectors
    calculated inside the subroutine {\tt DEKAY} placed in the file
    {\tt tauola.f}. The {\tt IFPSEUDO} is a logical parameter,
    defined in the subroutine {\tt TAUOLA} and placed in the file {\tt
    tauface-jetset.f}.  It tells  whether spin effects are calculated
    for the scalar or for the  mixed scalar--pseudoscalar Higgs boson. 
    If we have pure scalar Higgs boson it is set to
    {\tt .false.}. If we have mixed scalar--pseudoscalar case  it is set to {\tt
    .true.}. The pure pseudoscalar case is for $\phi= \frac{\pi}{2}$.
\item 
   We have assumed that  production generator provides two-body Higgs
   boson decays to $\tau$ leptons. In particular, that it  does
   not provide any bremsstrahlung corrections. Instead,  {\tt PHOTOS}
   \cite{Barberio:1990ms,Barberio:1994qi} (see also \cite{Andonov:2002mx}) can be used for that
   purpose, once the generation of $\tau^\pm$  decays is completed.
\item
   A more complete inclusion of  bremsstrahlung corrections would
   require a  substantial rewriting and extension of the program to the
   solution  presented in Ref. \cite{Jadach:1999vf} or a similar one.
\end{itemize}

%%%%%%%%%%%%%%%%%%%%%%%%%%%%%%%%%%%%%%%%%%%%%%%%%%%%%%%%%%%%%%%%%%%%%%%%%%%%
%%%%%%%%%%%%%%%%%%%%%%%%%%%%%%%%%%%%%%%%%%%%%%%%%%%%%%%%%%%%%%%%%%%%%%%%%%%%
\section{How to use the package}
%%%%%%%%%%%%%%%%%%%%%%%%%%%%%%%%%%%%%%%%%%%%%%%%%%%%%%%%%%%%%%%%%%%%%%%%%%%%
%%%%%%%%%%%%%%%%%%%%%%%%%%%%%%%%%%%%%%%%%%%%%%%%%%%%%%%%%%%%%%%%%%%%%%%%%%%%

\begin{enumerate}
\item
   Start with {\tt make Clean} from main directory  to secure against
   mismatches.
\item
   Check platform-dependent {\tt makefiles}.
\begin{itemize}
\item
    Go to subdirectory {\tt platform/}
\item
    Determine if {\tt make-xxx.inc} file specific for your computer is
present  there:  for Linux it is {\tt make-linux.inc}; for others it 
needs to be adapted.
\item
    Erase the symbolic link {\tt make.inc} which exists in this directory and
create a  new one, which points at the chosen {\tt make-xxx.inc}:
    \begin{itemize}
     \item {\tt rm make.inc}
     \item for Linux: {\tt ln -s make-linux.inc make.inc}
    \end{itemize}
   Afterwards check whether a link to the {\tt make.inc} is present in
   the  main directory.
\end{itemize}
\item
The settings of {\tt TAUOLA} input parameters  can be changed   for all
implemented applications,  see Section \ref{setting} for details.
\item
   {\tt PHOTOS} and {\tt TAUOLA} have their own private random
generators. If  you wish to replace them, you should do it at this
point, see Section \ref{randnumgen}.
\item
Create required versions of {\tt photos/} and {\tt tauola/}
directories. It  is mandatory to create {\tt photos/} directory first,
{\it i.e.} before creating {\tt tauola/}.
\begin{itemize}
\item
    Go to the directory {\tt photos-F}
\item
    Type one of the following commands to choose the required version
of {\tt  HEPEVT}:
\begin{itemize}
       \item
       {\tt make KK-all}
       \item
       {\tt make 2kD-all}
       \item
       {\tt make 4kD-all}
       \item
       {\tt make 2kR-all}
       \item
       {\tt make 10kD-all}
\end{itemize}
\item
   Go to directory tauola-F:
\item
   Type one of the following commands to choose the required version
of {\tt  TAUOLA} initialization:
\begin{itemize}
\item
      {\tt make cpc}
\item
      {\tt make cleo}
\item
      {\tt make aleph}
\end{itemize}
The user can then type
\begin{itemize}
\item
      {\tt make pythia}
\end{itemize}
to construct the additional demo executable with  {\tt PYTHIA}.
 In this demonstration program, old  {\tt PYTHIA} version 5.720 \cite{Sjostrand:1987hx}
is used. The user is anyway expected to attach the package to a new software,
up to date at the  time of installation. In this example, the matching
of the common block dimensions is not automatically assured. The option
 {\tt make 4kD-all} has to be used.
\end{itemize}
\item
The required version of {\tt PHOTOS} and {\tt TAUOLA} will reside in
newly  (re)created directories {\tt ./photos} and {\tt ./tauola}.
\item
The following {\tt demos} can be invoked from those directories:
\begin{itemize}
\item
   Demo for {\tt PHOTOS}  resides in {\tt ./photos/demo},  and can be
     invoked by command {\tt make} followed by {\tt make run}.
\item
   Demo for {\tt TAUOLA}  stand-alone resides in \\ {\tt
   ./tauola/demo-standalone},  and can be invoked by the command {\tt
   make} followed by  {\tt make run}.
\item
   Demo for {\tt TAUOLA} with  {\tt JETSET} being a host Monte Carlo
   resides in \\  {\tt ./tauola/demo-jetset}, and can be invoked by the
   command {\tt make} followed by  {\tt make run}.
\item
   Demo for {\tt TAUOLA} with  {\tt PYTHIA} being a host Monte Carlo
   resides in \\  {\tt ./tauola/demo-pythia}, and can be invoked by the
   command {\tt make} followed by  {\tt make run}.
\item
   Interface to {\tt KKMC}  resides in  {\tt
     ./tauola/KK-face/Tauface.f}. It has to be moved to  {\tt
     ./KK2f/Tauface.f}  of  distribution directory of  KKMC
     \cite{Jadach:1999vf}. The rest of the  {\tt ./tauola} directory
     should  replace the original one of the  {\tt KK} Monte Carlo
     distribution.
\end{itemize}
\end{enumerate}
Finally, let us remark that most of the {\tt TAUOLA} tree is not
necessary  and can be erased at this point.  Code and makefiles of
the directories {\tt ./tauola} and {\tt ./photos} are  sufficient.  To
execute demo programs,  the directories {\tt ./jetset} and {\tt
./glibk}  need to be kept or replaced by the appropriate links.  The
{\tt make.inc} symbolic link pointing to the {\tt make-xxxx.inc} file 
 in the directory  {\tt ./platform} also needs to be kept. The file {\tt make-xxxx.inc }
defines appropriate compiler flags.

%%%%%%%%%%%%%%%%%%%%%%%%%%%%%%%%%%%%%%%%%%%%%%%%%%%%%%%%%%%%%%%%%%%%%%%%%%%%
%%%%%%%%%%%%%%%%%%%%%%%%%%%%%%%%%%%%%%%%%%%%%%%%%%%%%%%%%%%%%%%%%%%%%%%%%%%%
\section{New hadronic currents for $\tau \to 4\pi$ 
channels of  $\tau$ decay}
%%%%%%%%%%%%%%%%%%%%%%%%%%%%%%%%%%%%%%%%%%%%%%%%%%%%%%%%%%%%%%%%%%%%%%%%%%%%
%%%%%%%%%%%%%%%%%%%%%%%%%%%%%%%%%%%%%%%%%%%%%%%%%%%%%%%%%%%%%%%%%%%%%%%%%%%%

The parametrization of formfactors developed for 4$\pi$ channels
of   $\tau$ decay and based on Novosibirsk data has been coded in a
form suitable for the {\tt TAUOLA} Monte Carlo 
package \cite{Bondar:2002mw}. There exist
two 4$\pi$ final  states in $\tau^{-}$ decay ( i.e.
$\tau^{-}\rightarrow \nu_{\tau}\pi^{-}\pi^{0}\pi^{0} \pi^{0}$,
$\tau^{-}\rightarrow \nu_{\tau}\pi^{-}\pi^{-}\pi^{+}\pi^{0}$).  Both
are available as parallel decay modes in the {\tt cpc}, {\tt cleo}
or  {\tt aleph} versions.  To get the new 4$\pi$ channels of $\tau$
decay ({\tt BINP} version) one should act in the  following way:
\begin{enumerate}
\item Go to directory {\tt tauola-factory}.
\begin{itemize}
\item There is a {\tt glue} programme that prepares a Fortran `precode' of
{\tt TAUOLA} library  and places it in the directory {\tt ../tauola-F}.
\item Type the following command to see the parameters of the {\tt
glue}  programme:
\begin{itemize}
\item  {\tt ./glue }
\end{itemize}
\item Type the following command to choose the {\tt BINP} version:
\begin{itemize}
\item  {\tt ./glue binp}
\end{itemize}
\item Type the following command to come back to the standard version:
\begin{itemize}
\item  {\tt ./glue standard}
\end{itemize}
\end{itemize}
\item Go to the main directory.
\item Act according to the instructions described in the previous
chapter.
\item The required {\tt TAUOLA} version ({\tt cpc}, {\tt cleo} or
{\tt aleph}) with new 4$\pi$ channels of $\tau$ decay ({\tt BINP}
version) will reside in newly (re)created directory {\tt ./tauola}.
\end{enumerate}

Another parametrization of form factors for 4$\pi$ channels
based on References \cite{Czyz:2000wh,Decker:1996af}  is available as well.
It can be installed, similar the case described above.
The only difference is that flag {\tt binp} has to be replaced 
with {\tt Karlsruhe}.

%%%%%%%%%%%%%%%%%%%%%%%%%%%%%%%%%%%%%%%%%%%%%%%%%%%%%%%%%%%%%%%%%%%%%%%%%%%%%%
%%%%%%%%%%%%%%%%%%%%%%%%%%%%%%%%%%%%%%%%%%%%%%%%%%%%%%%%%%%%%%%%%%%%%%%%%%%%%%
\section{Universal tool for HEP Monte Carlo generator comparison:
 {\tt MC-TESTER}}
%%%%%%%%%%%%%%%%%%%%%%%%%%%%%%%%%%%%%%%%%%%%%%%%%%%%%%%%%%%%%%%%%%%%%%%%%%%%
%%%%%%%%%%%%%%%%%%%%%%%%%%%%%%%%%%%%%%%%%%%%%%%%%%%%%%%%%%%%%%%%%%%%%%%%%%%%

Within the scope of the {\tt TAUOLA-PHOTOS} project work, a pilot project of
{\tt MC-TESTER} tool for semi-automatic comparisons of Monte Carlo generators
has been deployed. Initially, the project was targeted to be a part of this
package: it was intended to be used as a debugging/analysis tool for comparing 
the results produced by various versions of {\tt TAUOLA} and {\tt PHOTOS} code.

We decided however to publish {\tt MC-TESTER} as an independent 
project~\cite{Golonka:2002rz}. The scope of its possible use spans far 
beyond its original use case. We have however preserved the relationship
between the two projects: we include two versions of {\tt TAUOLA} code
(that may be produced using {\tt TAUOLA-PHOTOS} ) in the {\tt MC-TESTER} 
distribution (versions 1.0 and 1.1) as a test that verifies
the correctness of {\tt MC-TESTER} installation.

More information and the code of {\tt MC-TESTER} is available from its
web page  \cite{tester}.

%%%%%%%%%%%%%%%%%%%%%%%%%%%%%%%%%%%%%%%%%%%%%%%%%%%%%%%%%%%%%%%%%%%%%%%%%%%%%%
%%%%%%%%%%%%%%%%%%%%%%%%%%%%%%%%%%%%%%%%%%%%%%%%%%%%%%%%%%%%%%%%%%%%%%%%%%%%%%
\section{Minor upgrades in {\tt TAUOLA} and {\tt PHOTOS}}
%%%%%%%%%%%%%%%%%%%%%%%%%%%%%%%%%%%%%%%%%%%%%%%%%%%%%%%%%%%%%%%%%%%%%%%%%%%%
%%%%%%%%%%%%%%%%%%%%%%%%%%%%%%%%%%%%%%%%%%%%%%%%%%%%%%%%%%%%%%%%%%%%%%%%%%%%
There is no need to publish the new versions of the programs: 
 since the last publication, the upgrades are minor.
Nonetheless let us list here the main improvements collected over the years.
The list of fixes for  numerically rather insignificant bugs, as well as necessary
changes due to the evolution of compilers, will not be given here.  
For that purpose we refer the reader
to the comments in the code, to the {\tt README} files included in the 
distribution and to the package web page \cite{wasm}.

\vskip 1 mm
\subsection{{\tt TAUOLA}}
\vskip 1 mm

Apart from the modifications listed in Section 1, no changes were
introduced in the {\tt cpc} and {\tt aleph} initializations of {\tt TAUOLA}
 since the time they became available. 
In the case of {\tt cleo}  initialization,
a change was introduced in the $\tau \to \nu \pi^+ \pi^+ \pi^- \pi^0$ channel;
normalization of the $\omega$
current
was adjusted to experimental data \cite{Hagiwara:2002fs}. 
To this end, the $\omega$  contribution was diminished from 68$\%$
down to 40$\%$ of the total rate for the  $\tau \to \nu \pi^+ \pi^+ \pi^- \pi^0$
channel.  
The  new version of {\tt TAUOLA} 2.7, was nonetheless introduced, which
is marked in all initialization printouts, to document progress in bug fixing etc. 
The creation dates printed in the outputs are retained.
They  mark the time when  the bulk of the work was done.

Since publication~\cite{Bondar:2002mw} a careful analysis of a hadronic
current for $4\pi$ final states was performed; related extensive studies
and comparisons are documented in Ref.~\cite{TomekPhD}. The new
parametrization based on Refs.~\cite{Czyz:2000wh,Decker:1996af} was also 
installed in {\tt TAUOLA} . Different options of models 
discussed in Ref.~\cite{Bondar:2002mw}  were probed.
   Let us list changes with respect to ~\cite{Bondar:2002mw},
which are now installed in the {\tt TAUOLA} code 
if used with the {\tt binp} option.  
We will use the opportunity to list misprints found in Ref.~\cite{Bondar:2002mw} as well:

\begin{enumerate}
\item
In Table 4 of Ref.\cite{Bondar:2002mw} in all columns of function arguments,
 $\sqrt{Q^2}$ were listed  rather than  
 $Q^2$, as written in the table header.
\item
Formula (20) of \cite{Bondar:2002mw} should read $g_\sigma (s)=(1-4m^2/s)^{1/2}$, rather than  
$g_\sigma (s)=(s-4m^2/s)^{1/2}$.
\item
All propagators derived from  formula (18) of Ref.~\cite{Bondar:2002mw}, should be normalized to $-1$ 
at $q^2=0$;
also imaginary part should be set to zero below opening of the appropriate 
decay channel (i.e. typically for $q^2<4m^2$).
\item
In the definition of the $\rho$ propagator in formula (18) of Ref.~\cite{Bondar:2002mw}, the term
$dm(q^2)$ was neglected. The full $\rho$ propagator should
read
%$D(q)=q^2-M^2 + iM\Gamma\frac{g(q^2)}{g(M^2)}$, to 
\begin{eqnarray}
D_\rho(q^2)&=&q^2-M_\rho^2 -M_\rho \Gamma _\rho ~dm(q^2) + iM_\rho\Gamma_\rho\frac{g_\rho(q^2)}{g_\rho(M^2_\rho)} \nonumber \\
dm(q^2)&=&\Big(h_\rho(q^2)-h_\rho(M^2_\rho)-(q^2-M_\rho^2)
\frac{dh_\rho(q^2)}{ dq^2}\Big|_{q^2=M^2_\rho}
\Big)/g_\rho(M^2_\rho),
\end{eqnarray}
where
\begin{equation}
h_\rho(q^2) = \left\{
 \begin{array}{cl}	
\Huge{\frac{\sqrt{1-\frac{4m^2}{q^2}}\ln\Big(\frac{1+\sqrt{1-\frac{4m^2}{q^2}}}{1-\sqrt{1-\frac{4m^2}{q^2}}}\Big)(q^2-4m^2)}{\pi}} & {\rm ~for~} q^2 > 4m^2, \\ \\
 -8 \frac{m^2}{\pi} & {\rm ~for~} q^2 = 0 {\rm ~GeV}^2, \\ \\
 0 & {\rm ~~otherwise~}.
\end{array}\right.
\end{equation}
\item
The full $\omega $ propagator should read (the $\frac{g_\omega(q^2)}{g_\omega(M^2_\omega)} $ was 
approximated by 1):
\begin{eqnarray}
D_\omega(q^2)&=&q^2-M_\omega^2  + iM_\omega\Gamma_\omega\frac{g_\omega(q^2)}{g_\omega(M^2_\omega)} \nonumber \\
%& & \nonumber \\
\end{eqnarray}
%\begin{displaymath}
\begin{equation}\label{g_omega}
\hspace{-2.cm} \frac{g_\omega(q^2)}{g_\omega(M^2_\omega)}  = \left\{ 
  \begin{array}{cl}
%   \: 
  {\rm max} \left( 0~, \begin{array}{l} 
1+17.560(\sqrt{q^2}-M_\omega)+141.110(\sqrt{q^2}-M_\omega)^2+\\+894.884(\sqrt{q^2}-M_\omega)^3 +4977.35(\sqrt{q^2}-M_\omega)^4+\\+7610.66(\sqrt{q^2}-M_\omega)^5-42524.4(\sqrt{q^2}-M_\omega)^6
	\end{array} \right)&  {\rm ~~for~}  \sqrt{q^2}<1.0 {\rm ~GeV},
%0.635329 
%	\hspace{1cm}  
\\ \\
  -1333.26 + 4860.19\sqrt{q^2} -6000.81(\sqrt{q^2})^2 +2504.97(\sqrt{q^2})^3& 
{\rm ~~otherwise~} .
  \end{array}
  \right.
%\end{displaymath}
\end{equation}
Function (\ref{g_omega})
is involved, because of effects due to three-body phase space, 
threshold effects of pion masses as well as opening of the resonant $\rho$
channel in $\omega$ decay.
\item
In formula (16) of Ref.~\cite{Bondar:2002mw} the overall sign  was missing.
\item
In formula (21) of Ref.~\cite{Bondar:2002mw}, the coefficient in front of the second line should include $s$ instead of
$a$.
%\item
%in Formula (21)~\cite{Bondar:2002mw} an argument of $a_1$ form factor $F_{a_1}$ should read $q^2$.
\item\label{ga1}
We have recalculated, using two methods, the outcome 
from formula (21) of \cite{Bondar:2002mw}. We have found some numerical difference with respect 
to the one used  before. The size of the difference is small, except
the region above 1.3~GeV, which is not important anyway (see   Ref.~\cite{TomekPhD}). 
\item
 In the comment for formula (21) of Ref.~\cite{Bondar:2002mw} it should be stated that 
three-vectors and energies are defined in the $a_1$  (3$\pi$) system rather than in the $\tilde \rho$
(4$\pi$) system. 
\item
In formula (25) of Ref.~\cite{Bondar:2002mw}, the overall coefficient  $z_{mix}$ was missing, but it was
set in the code to $z_{mix}=1$ anyway.
\item
 At the beginning of Section 5  in Ref.~\cite{Bondar:2002mw} the numerical constant 
 $\Lambda^2= 1.2$ GeV
should be defined, rather than  $\Lambda$ alone.
\end{enumerate}

All points from the above list do not change the numerical content
of the code, except for point \ref{ga1}, which brings a modification, and consequently 
leads to numerically small compensating changes of the content  
of the tables of functions $G$ presented in Ref. \cite{Bondar:2002mw}.

We found,  while playing with different versions 
of the  {\tt TAUOLA} code describing the  
Novosibirsk hadronic current, that we had for some time
an overall sign error in the space-like component of the 
current.
For all distributions involving pions only, it was fully compensated 
by the choice of the fit functions $G(Q^2)$ 
(formulae (14), (15) and (22--24) of Ref. \cite{Bondar:2002mw}). Also
all  figures in Ref.~\cite{Bondar:2002mw} are correct.

However, as a consequence of the bug, the distributions involving 
$\nu_\tau$  were affected in {\tt TAUOLA BINP}.
The effect was not large enough for
comparisons with {\tt TAUOLA CLEO} to hint for
the problem. The differences were within the expected systematic errors of the two 
parametrizations. 
We have also found an additional misprint
 in the numerical value for our fit functions: 
$G_{\pi^+\pi^-\pi^+\pi^0}$, $G_{\pi^+\pi^-\pi^+\pi^0}^\omega $ and
$G_{\pi^+\pi^0\pi^0\pi^0}$  (see Table 4, Ref. \cite{Bondar:2002mw}).  
An overall constant or  function was missing.
Once corrections  are taken together, the 
values in Table 4 should be multiplied respectively by:
\begin{itemize}
\item
\begin{equation}\label{G1}
\frac{76.565\, \sqrt{0.71709*\sqrt{Q^2} - 0.27505}}{M_\rho^4\sqrt{Q^2}}~{\rm ~~~for~ the~~} G_{\pi^+\pi^-\pi^+\pi^0}(Q^2),
\end{equation}
\item
\begin{equation}\label{G1_omega}
\frac{886.84\,\sqrt{0.70983*\sqrt{Q^2} - 0.26689}}{M_\rho^4\sqrt{Q^2}}~{\rm ~~~for~ the~~} 
G_{\pi^+\pi^-\pi^+\pi^0}^\omega(Q^2) ,
\end{equation}
\item
\begin{equation}\label{G2}
\frac{96.867\,|{\tt z\_forma1(Q^2)}|\sqrt{0.70907*\sqrt{Q^2} - 0.26413}}{M_\rho^4\sqrt{Q^2}}~ {\rm ~~~
for~ the~~} G_{\pi^+\pi^0\pi^0\pi^0}(Q^2)
\end{equation}
\end{itemize}
before being used in formulae (14), (15) and (22--24) of Ref. \cite{Bondar:2002mw}.
%The ${\tt z\_forma1(Q^2)}$ complex function describes the 
%energy dependence of the cross section $e^+e^- \to a_1 \pi \to \pi^+\pi^-\pi^+\pi^-$
%(see paragraph A.1 of  Ref. \cite{Akhmetshin:df}).
Numerical values of the ${\tt |z\_forma1(Q^2)|}$ function are tabulated in 
the new
{\tt TAUOLA} function
{\tt ZFA1TAB(Q2)}, as was  the case with the functions $G$.

Finally, during the work, as a byproduct, we have switched on the interference between
the $\omega$ and non-$\omega$ parts of this current. This changed the  overall rate,
mainly because of the interference of small, spread over all phase space
tails of the $\omega$ Breit-Wigner propagator. The numerical effect is not visible in
the differential distribution, but it affects the total rate by 1.017(1).
 We changed
the normalization of $G_{\pi^+\pi^-\pi^+\pi^0}(Q^2)$ and
$G_{\pi^+\pi^-\pi^+\pi^0}^\omega(Q^2)$ by the same factor,
%non-$\omega$-resonant part of the current,}
so as to reproduce the total 4$\pi$ rate as in PDG 2000.
Note that the interference of the $\omega$ current with the remaining part 
provides 1.7 \% effect on the total rate. 

\vskip 1 mm
\subsection{{\tt PHOTOS }}
\vskip 1 mm
The changes we introduced are collected in the following subsections.

\vskip 1 mm
\subsubsection{{\tt Subroutine {\tt PHCORK}} }
\vskip 1 mm
% CorrectBranchKinematics() [PHCORK in F77 version]
During the test phase of the {\tt Photos+}  \cite{MsCGolonka}, a C++
 implementation of the  {\tt PHOTOS} algorithm, it turned
out that the generation becomes unstable for events produced  
 at TeV energies. The use of {\tt PHOTOS} at
this energy range was then studied for the first time.
The kinematics of events produced by  {\tt PYTHIA} 5.7 
was often not precise enough, because of rounding errors for
({\tt REAL*4}) floating-point variables.

The particles described by a
tree-structured event record (i.e. {\tt HEPEVT} common block used by {\tt PHOTOS}),
form decay `branches' composed of a `mother' -- the decaying particle,
 and `daughters' -- its decay products. 
When the kinematics of a branch is not correct
(the sum of 4-momentum  of decay products does not match 
the 4-momentum of the mother particle or the particles momenta are off shell), 
numerical instabilities are
encountered. It is not possible to perform boosts of the particles 
correctly if the boost parameters are large (in certain of the events
where errors occurred, we found, for instance, a pion with momentum of order
of TeV/c). 

A special routine {\tt PHCORK}, which corrects these kinds of inconsistencies,
have been developed and inserted to {\tt PHOTOS} code\footnote{This routine became available 
in 1999 and can be found in the {\tt CERNLIB} version of {\tt PHOTOS}.}.

The routine works in one of four modes, selected at the initialization of 
{\tt PHOTOS}:

\begin{itemize}
        \item{mode 1}: no correction performed,
        \item{mode 2}: energy is corrected from mass,
        \item{mode 3}: mass is corrected from energy,
        \item{mode 4}: energy is corrected from mass for particles up to a mass of 0.4~GeV,
for heavier ones: mass is corrected.
\end{itemize}

In cases, where a branch (sub-cascade starting from certain particles) has two
mothers, the first mother's momentum is corrected according to the sum of its children
momenta and the momentum of the second mother.

The default mode  is 1, which means that no
correction is applied, and  full compatibility with older implementations is retained.

Correction is performed in two steps. First, all daughters' energies 
(or masses, depending on the requested mode of correction) 
are corrected to fulfil the 
       $ E^2  - p^2 = M^2$ constraint.

In the second step, the sum of all daughters' 4-momenta  is calculated.  The
mother's momentum is corrected to this  value. In cases where two
mothers are present, only the first mother's 4-momentum is
corrected to a value:

      $  P_{mother_1} = ( \sum _{daughters} P_{daughter}) - P_{mother_2}$.

\vskip 1 mm
\subsubsection{Interference correction}
\vskip 1 mm
   In the published version of {\tt PHOTOS} \cite{Barberio:1994qi} the interference 
between photon emission from two sources was available only for two-body decays
into  particles of opposite charge and equal mass.
Since then, the interference correction was extended to work also in the case of decays
into particles of distinct masses.
On the technical side the interference correction was re-coded to work more efficiently. 
 For example this should prevent the program to stop
in case of  double bremsstrahlung and interference 
corrections in the decay of the very heavy $Z/\gamma^* \to e^+e^-$ state. 

\vskip 1 mm
\subsubsection{Enabling flags}
\vskip 1 mm

For some users, it is convenient to block/enable {\tt PHOTOS} generation in some  cases,
for tests or because of unphysical coding of events into the {\tt HEPEVT} common block.
Flags to block/enable {\tt PHOTOS} generation in 
$\pi^0 \to e^+e^- \gamma$ and in $W$ decay into quarks were  introduced. 
The flags  are initialized in the subroutine {\tt  PHOINI}.

\vskip 1 mm
\subsubsection{Weight correction in the decay $W\to l \nu_l$}
\vskip 1 mm

For  the leptonic decays of the $W$ boson, {\tt PHOTOS}
predictions were carefully  compared with the results from the first order
matrix element generator \cite{Andonov:2002mx}. It was found that even though the
leading corrections were properly modelled by  {\tt PHOTOS},
the non-leading missing effects were relatively large.
The origin of these discrepancies 
was  understood, and  the appropriate correcting weight was introduced into 
{\tt PHOTOS}. With these changes the agreement with the matrix element calculation 
 improved sizeably \cite{Nanava:2003cg}, especially in regions of phase space  populated
 with hard non-collinear photons.
A new option was introduced, to initialize this correcting weight.
It can be switched on/off  with the new flag {\tt IFW},
initialized in subroutine {\tt PHOCIN}.
The method can easily be extended to other processes
where the matrix element is available  from QED calculations or from modelling
based on experimental data. 

\vskip 1 mm
\subsubsection{{\tt PHOTOS} 2.07 }
\vskip 1 mm

Once all these modifications are installed, the new version 2.07 of {\tt PHOTOS} was
created. 
 
%%%%%%%%%%%%%%%%%%%%%%%%%%%%%%%%%%%%%%%%%%%%%%%%%%%%%%%%%%%%%%%%%%%%%%%%%%%%%%
%%%%%%%%%%%%%%%%%%%%%%%%%%%%%%%%%%%%%%%%%%%%%%%%%%%%%%%%%%%%%%%%%%%%%%%%%%%%%%
\section{Issues related to distinct event record schemes}
%%%%%%%%%%%%%%%%%%%%%%%%%%%%%%%%%%%%%%%%%%%%%%%%%%%%%%%%%%%%%%%%%%%%%%%%%%%%
%%%%%%%%%%%%%%%%%%%%%%%%%%%%%%%%%%%%%%%%%%%%%%%%%%%%%%%%%%%%%%%%%%%%%%%%%%%%
\label{sekcja8}

Let us note that the packages presented here work with the standard {\tt HEPEVT}
event record, where the tree structure of the event evolution is properly coded
into a unique tree structure. Generally this is not the case. However, the programs
were shown to work properly with the extension of the standard of the type as
originating from {\tt PYTHIA} \cite{Sjostrand:2000wi} with {\tt
MSTP(128)=0,1,2}.  Necessary adjustments turned out to be non-trivial and in
fact correctness of their action cannot be garanteed in the general case
\cite{july2003}.  The case of {\tt HERWIG} \cite{HERWIG} turned out to be even
more complex. Even though in this case the {\tt HEPEVT} common block is used as
standard event record, the mother--daugter pointers are nonetheless used in a
different way. As a consequence, events generated by {\tt HERWIG} are far from
the design format of the {\tt HEPEVT} event record.  We were able to tune the 
working of the {\tt TAUOLA universal interface} to the requirements of operating
with {\tt HERWIG}; however, the interfacing of {\tt PHOTOS} turned out to be
rather more difficult. 

\begin{figure}
\begin{center}
\epsfig{file=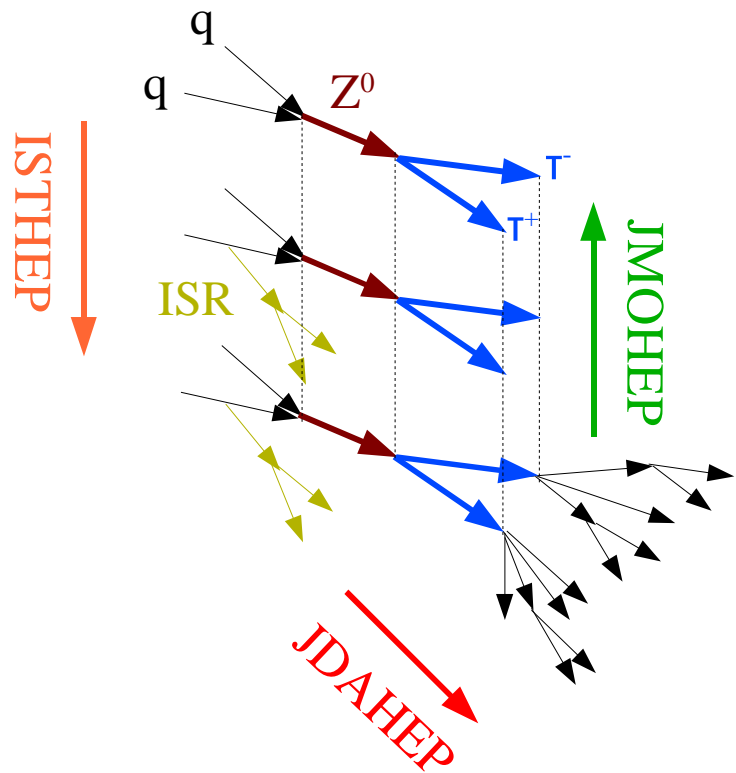,width=0.6\linewidth}
\end{center}
\caption{The usage of {\tt JMOHEP} and {\tt JDAHEP} pointers in {\tt PYTHIA} with {\tt
MSTP(128)=0} setting. While {\tt PYTHIA} keeps the {\tt JDAHEP} entries to point
in the direction of event evolution, the {\tt JMOHEP} is used to point to the
copies of a particle produced when perturbative effects, such as initial state
radiation(ISR) or $\tau$ decays are added; the particle copies at different
stages of event simulation are consistently distinguished by the {\tt ISTHEP} tags. 
\label{fig:HEPEVT_py}}
\end{figure}

The principal issue from which difficulties with the {\tt HEPEVT} decoding arise
is that {\tt PYTHIA} as well as {\tt HERWIG} produce a new copy of the
participating particles each time a physical effect (e.g. final state
radiation) is added; consequently several copies of particle entries are
produced, which cannot be consistently put in the {\tt HEPEVT} structure. In
order to circumvent this problem, the {\tt JMOHEP} and/or {\tt JDAHEP} entries
are {\it overloaded}, i.e. the information stored in them is used for other
purposes than their original one. For example, {\tt PYTHIA} with {\tt
MSTP(128)=0} setting uses the {\tt JMOHEP} entry to point to the copies of a
particle (cf. Fig. \ref{fig:HEPEVT_py}) and {\tt HERWIG} uses the {\tt JMOHEP(2,I)} entries to store the colour
information of the particle flow. A generic interface thus has to be informed of
these possibilities and to correct for them for the {\tt TAUOLA} and {\tt
PHOTOS} procedures to work properly.

At present a working version of modified {\tt PHOTOS} routines, which appears to
work with {\tt HERWIG}, can be found in the distribution of {\tt AcerMC}
\cite{AcerMC}. However rigorous testing is necessary before these modifications
can be included in the official {\tt PHOTOS} distribution. The original {\tt PHOTOS}  code
had to be modified due to the `overloaded' {\tt HEPEVT} record, since its
requirements for matching mother--daughter pointers were too strict for either
{\tt PYTHIA} or {\tt HERWIG}. The modification was limited to {\tt PHOTOS\_MAKE}
and {\tt PHOBOS} routines. In addition the tracking of the {\tt IDHW} array was
added to {\tt PHOTOS\_MAKE} to accommodate the {\tt HERWIG} event record. A further
modification was however necessary in the {\tt PHOIN} routine since in {\tt
HERWIG} the entry {\tt JMOHEP(2,I)} is not empty but filled with colour flow
information, which in turn inhibited {\tt PHOTOS} radiation off participating
particles\footnote{{\tt PHOTOS} expects the non-zero second mother {\tt
JMOHEP(2,I)} entry only for $gg(qq) \to t\bar{t}$ process, which is treated by a
set of dedicated routines; this in turn clashes with {\tt HERWIG} overloaded {\tt
JMOHEP(2,I)} entry.}

It has to be stressed that implementing the {\tt HERWIG} interface to {\tt
PHOTOS} is desirable since up to version 6.5 {\tt HERWIG} did not
have any QED radiation implemented in the decays. In the later releases,
{\tt PHOTOS} is installed in decays (and hadronic production) of $t$-quarks,
as well as in $W$ decays. This is done, however,   
through specifically tailored interfaces embodied within the {\tt HERWIG}  code.

Another two issues are related to the double counting. If there is no
information passed from the host program and the common block {\tt PHOQED} of
{\tt PHOTOS} is not appropriately filled, then there is no way for {\tt PHOTOS}
to know that in some decays it should not generate bremsstrahlung (again) and
thus `double counting' will appear. While the question is not relevant for {\tt
HERWIG}, since no photon bremsstrahlung is implemented, it can become an issue
in {\tt PYTHIA}, for some specific processes. Indeed, one can generally
switch off the photon radiation off leptons (by setting a high limit on {\tt
PARJ(90)}, cf. \cite{Sjostrand:2000wi}); however, one cannot do it on the basis of specific
processes and/or decays. 

The issue is not shared in the {\tt TAUOLA universal interface}, since the 
decayed $\tau$ leptons are not decayed again and the original decay
will remain.

%%%%%%%%%%%%%%%%%%%%%%%%%%%%%%%%%%%%%%%%%%%%%%%%%%%%%%%%%%%%%%%%%%%%%%%%%%%%%%
%%%%%%%%%%%%%%%%%%%%%%%%%%%%%%%%%%%%%%%%%%%%%%%%%%%%%%%%%%%%%%%%%%%%%%%%%%%%%%
\section{Summary and future possibilities}
%%%%%%%%%%%%%%%%%%%%%%%%%%%%%%%%%%%%%%%%%%%%%%%%%%%%%%%%%%%%%%%%%%%%%%%%%%%%
%%%%%%%%%%%%%%%%%%%%%%%%%%%%%%%%%%%%%%%%%%%%%%%%%%%%%%%%%%%%%%%%%%%%%%%%%%%%

We have presented the system for creating the required version of {\tt
PHOTOS} and {\tt TAUOLA} packages from their master versions. The
master versions are structured  in a relatively compact form without
code duplications etc. The minor modifications of the programs,
introduced since their publication, were also explained.

The creation of the system  
was the first step towards future attempts to develop  packages
without loss of their present physics contents.  Some experience,
already collected in that direction, is summarized in
\cite{MsCGolonka}.  We find the question of the language translation
for  the fixed program version relatively easy. On the contrary, the question
of project continuity into further upgrades motivated by the physics
needs to be thought over carefully. Matching the  programming styles of
the C++ experts with the strategies of testing numerical
correctness of consecutive versions is a rather crucial issue, which
has to be addressed.  Tools and methods embodied  in Fortran code survive
such translation with difficulty.

At a certain moment, the necessary strategy may thus require fluency  in the
Fortran, Object-Orientated languages and physics content of the project by the
same person. Platform-independent  tools for mixing code in Fortran
and OO languages might be of great help.

In this paper we have also documented for the first time technical side
of the universal interface for the {\tt TAUOLA} package. From now  on, its version number 
1.10 is printed in the output.

%%%%%%%%%%%%%%%%%%%%%%%%%%%%%%%%%%%%%%%%%%%%%%%%%%%%%%%%%%%%%%%%%%%%%%%%%%%%
%%%%%%%%%%%%%%%%%%%%%%%%%%%%%%%%%%%%%%%%%%%%%%%%%%%%%%%%%%%%%%%%%%%%%%%%%%%%
\section*{Acknowledgements}
%%%%%%%%%%%%%%%%%%%%%%%%%%%%%%%%%%%%%%%%%%%%%%%%%%%%%%%%%%%%%%%%%%%%%%%%%%%%
%%%%%%%%%%%%%%%%%%%%%%%%%%%%%%%%%%%%%%%%%%%%%%%%%%%%%%%%%%%%%%%%%%%%%%%%%%%%

Authors are grateful to ALEPH, CLEO, DELPHI and OPAL collaborations
for providing their appropriate versions of the whole or parts of
TAUOLA  initialization. Useful discussions and suggestions from
B. Bloch, C. Biscarat, S. Jadach, J. H. K\"uhn, M. Peskin, S. Slabospitzky 
and  A. Weinstein are also acknowledged.

%Two of the authors (T.P. and M.W.) are grateful for the warm
%hospitality  extended to them by  the Budker Institute of Nuclear
%Physics in Novosibirsk. They also would  like to
%thank  the ``Marie Curie Programme'' of the European Commission for 
%fellowships.

%%%%%%%%%%%%%%%%%%%%%%%%%%%%%%%%%%%%%%%%%%%%%%%%%%%%%%%%%%%%%%%%%%%%%%%%%%%%
%%%%%%%%%%%%%%%%%%%%%%%%%%%%%%%%%%%%%%%%%%%%%%%%%%%%%%%%%%%%%%%%%%%%%%%%%%%%
%\bibliographystyle{utphys_spires}
%\bibliographystyle{plain}
%\bibliography{TAUOLA-F}

\providecommand{\href}[2]{#2}

\end{document}